\documentclass[twocolumn,showkeys,preprintnumbers,amsmath,amssymb]{revtex4}

\usepackage{graphicx}
\usepackage{dcolumn}
\usepackage{bm}

\begin{document}


\title{The Cosmic Time in Terms of the Redshift}

\author{Moshe Carmeli}
 \email{carmelim@bgu.ac.il}
\affiliation{Department of Physics, Ben Gurion University of the Negev, Beer Sheva
84105, Israel}%

\author{John G. Hartnett}
 \email{john@physics.uwa.edu.au}
\affiliation{School of Physics, the University of Western Australia\\35 Stirling Hwy, Crawley 6009 WA Australia}%

\author{Firmin J. Oliveira}
 \email{f.oliveira@jach.hawaii.edu}
\affiliation{Joint Astronomy Centre, Hilo, Hawai'i, U.S.A.}

\date{\today}

\begin{abstract}
In cosmology one labels the time $t$ since the Big Bang in terms of the 
redshift of light emitted at $t$, as we see it now. In this Note we derive a
formula that relates $t$ to $z$ which is valid for all redshifts.
One can go back in time as far as one wishes, but not to the Big Bang at
which the redshift tends to infinity.
\end{abstract} 

\maketitle

\section{Introduction}
In a recent paper by Renyue Cen and Jeremiah P. Ostriker \cite{Cen1999}, and in the review by Bertram Schwarzschild \cite{Schwarzschild2005}, the important problem of the missing 
baryons was discussed. That is, the fraction of baryons observed at high
redshift is about twice that observed at low redshift. Cen and Ostriker
performed a hydrodynamic computer simulation of cosmic evolution from about 2
billion years after the Big Bang to the present. The precise time scale 
depends on the details of the cosmological model from which it is derived. 
According to Schwarzschild \cite{Schwarzschild2005} cosmologists relate the cosmic time $t$ to 
cosmological redshift $z$ by roughly $t\approx 14\mbox{\rm Gyr}/(1+z)^{3/2}$. 
Here $t$ denotes the time since the Big Bang, that depends on the redshift $z$
of light emitted at time $t$, as we see it now.

The Cen-Ostriker simulation covered the interval from $z=0$ to 3. In the 
simulation, the intergalactic gas gets steadily hotter after $z=3$. More and 
more of it is shock heated as it repeatedly falls into gravitational potential  
wells of accumulated nonbaryonic ``dark" matter. As the gas gets hotter, less 
and less of it remains un-ionized, so that Lyman $\alpha$ absorption lines 
become increasingly harder to see. They suggest that as a result we cannot see
about 50\% of the baryonic matter. Cen and Ostriker conclude that the missing
baryons reside in a `warm-hot intergalactic medium'. It's clear, however, that 
for further investigation of the problem one needs a more accurate formula for 
the relationship between $t$ and $z$ that is valid for all redshift values. 
This may allow simulations of this kind closer to the time of the Big Bang.

\begin{figure}
\includegraphics[width = 3.5 in]{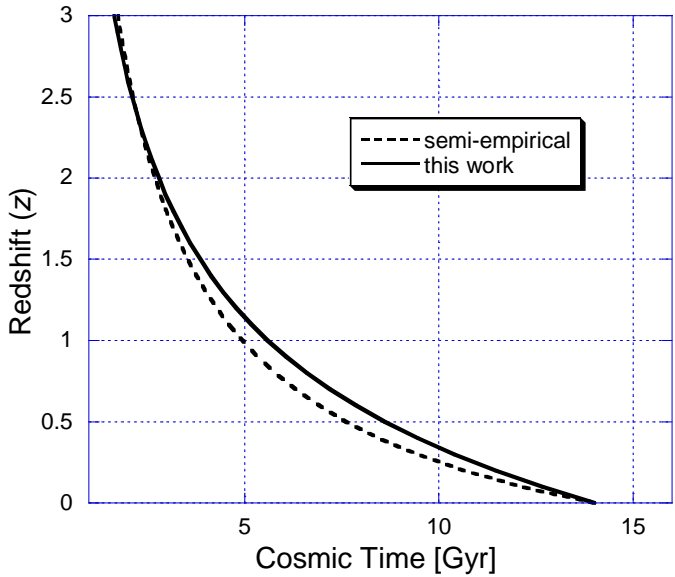}
\caption{\label{fig:fig1} Cosmic time as a function of redshift ($z$) where the solid line
represents Eq. (\ref{eqn:newformula}) (which is identical with Eq. (\ref{eqn:newredshiftformula5})) derived in this paper and the broken line represents 
$t\approx 14\mbox{\rm Gyr}/(1+z)^{3/2}$ for the semi-empirical model. The
equation  (\ref{eqn:newformula}) is valid over all $z$ and is further shown in Figs. \ref{fig:fig2} and \ref{fig:fig3}.}
\end{figure}

In this Note we derive such a formula: 
\begin{equation} \label{eqn:oldformula}
t=\frac{2H_0^{-1}}{1+\left(1+z\right)^2}
\end{equation}
where $H_0$ is the Hubble parameter.
For an appropriate choice of $H_0$ (70 km/s-Mpc), we obtain 
\begin{equation} \label{eqn:newformula}
t \approx \frac{28}{1+\left(1+z\right)^2}\mbox{\rm Gyr}.
\end{equation}
The formula is valid for all $z$. The dependence on $z$ is essentially 
different from that in the simulated formula. However, for $z$ from 0
to 3 (see Fig. \ref{fig:fig1}) the values of $t$ in both cases are very close and their 
difference essentially negligible. 
\section{Cosmological Line Element}
The Universe expands, of course, by the Hubble law $x=H_0^{-1}v$. But one 
cannot use this law directly to obtain a relation between $z$
and $t$. So we start by assuming that the Universe is empty of gravitation.
One can then describe the property of expansion as a null-vector in the flat 
four dimensions of space and expanding velocity $v$ \cite{Carmeli1995, Carmeli1996, Carmeli2002}.

The cosmological line element \cite{Carmeli1995, Carmeli1996} is given by 
$ds^2=\tau^2dv^2-(dx^2+dy^2+dz^2)$, where $\tau$ is the Big Bang time, the 
reciprocal 
of the Hubble parameter $H_0$ in the limit of zero distances, and it is a 
constant in this epoch of time. Its value is given by $\tau=12.486$ Gyr 
(Carmeli \cite{Carmeli2002}, p.138, Eq. (A.66)). When $ds=0$ one gets the Hubble expansion 
with no gravity. This line element should be compared with the Minkowskian 
line element in standard relativity,
$ds^2=c^2dt^2-(dx^2+dy^2+dz^2)$. When $ds=0$ in the latter case, one gets the 
equation of motion for the propagation of light. 

Space and time coordinates transform according to the Lorentz transformation,
\begin{subequations}
\begin{eqnarray}
x'=\left(x-vt\right)/\left(1-v^2/c^2\right)^{1/2}, \label{eqn:Lorentz1}
\\
t'=\left(t-vx/c^2\right)/\left(1-v^2/c^2\right)^{1/2}, \label{eqn:Lorentz2}
\end{eqnarray}
\end{subequations}
in ordinary physics. In cosmology the coordinates transform by the 
cosmological transformation (See \cite{Carmeli1995, Carmeli1996} and Sect. 2.11, p. 15 of \cite{Carmeli2002}),
\begin{subequations}
\begin{eqnarray}
x'=\left(x-tv\right)/\left(1-t^2/\tau^2\right)^{1/2}, \label{eqn:CSR1}
\\
v'=\left(v-xt/\tau^2\right)/\left(1-t^2/\tau^2\right)^{1/2}.\label{eqn:CSR2}
\end{eqnarray}
\end{subequations}
where $t$ is the cosmic time with respect to us now. 

Comparing the above 
transformations shows that the cosmological one can formally be obtained from
the Lorentz transformation by changing $t$ to $v$ and $c$ to $\tau$ ($v/c
\rightarrow t/\tau$). Thus the transfer from 
ordinary physics to the expanding Universe, under the above assumption of 
empty space, for null four-vectors is simply achieved by replacing $v/c$ by 
$t/\tau$, where $t$ is the cosmic time measured with respect to us now. 

Thus classical physical laws in which velocities appear can be transferred to
cosmology by replacing the velocity by the cosmic time measured with respect 
to us now ($v/c\rightarrow t/\tau$). For example, one can use the apparatus of 
four dimensions ($ct$, $x$, $y$, $z$) well known in electrodynamics. Using the
wave four-vector ($\omega$, {\bf k}) one can easily derive the transformation
of $\omega$ and {\bf k} from one coordinate system to another. This then gives 
the Doppler effect. 

A charged particle 
receding from the observer with a velocity $v$ and emitting electromagnetic 
waves will experience a frequency shift given by (see, for example, L. Landau 
and E. Lifshitz, p.121 \cite{Landau1959})
\begin{equation} \label{eqn:Dopplerformula}
\omega=\omega'\sqrt{\frac{1+v/c}{1-v/c}},
\end{equation}
where $\omega'$ and $\omega$ are the frequencies of the emitted radiation
received from the particle at velocity $v$ and at rest, respectively. And thus 
a redshift is obtained from
\begin{equation} \label{eqn:redshiftformula}
1+z =\sqrt{\frac{1+v/c}{1-v/c}}.
\end{equation}
\begin{figure}
\includegraphics[width = 3.5 in]{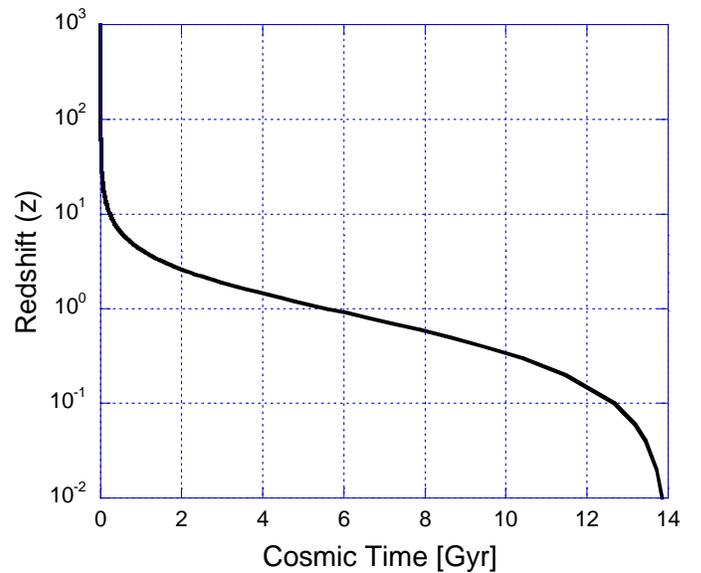}
\caption{\label{fig:fig2} Cosmic time as a function of redshift $z$ from Eq. (\ref{eqn:newredshiftformula5}) with time 
measured since the Big Bang.}
\end{figure}
\begin{figure}
\includegraphics[width = 3.5 in]{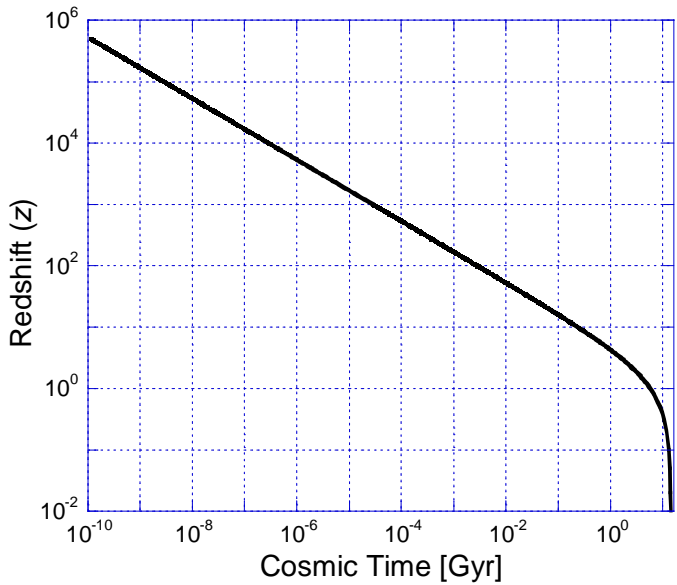}
\caption{\label{fig:fig3} Cosmic time as a function of redshift $z$ from Eq. (\ref{eqn:newredshiftformula5}) in a 
logarithmic scale.}
\end{figure}

In our case $\tau$ replaces $c$ and $t$ replaces $v$ ($v/c$ goes over to 
$t/\tau$), thus getting 
\begin{equation} \label{eqn:newredshiftformula}
1+z =\sqrt{\frac{1+t/\tau}{1-t/\tau}}.
\end{equation}
Rearranging we get
\begin{equation} \label{eqn:newredshiftformula2}
\frac{t}{\tau}=\frac{(1+z)^{2}-1}{(1+z)^{2}+1}.
\end{equation}

Now because the time $t$ is measured backwards, we need to make the 
substitutions such that $t$ is measured forward from the Big Bang, i.e., 
$t\rightarrow\tau-t$. Hence we obtain
\begin{equation} \label{eqn:newredshiftformula3}
t=\frac{2\tau}{1+\left(1+z\right)^2}.
\end{equation}
where $t$ now denotes the time since the Big Bang. 

The above formula was derived for the flat space. In order to extend the
result to the present state where Hubble's law determines the expansion,
we adopt the method used in classical general relativity theory. In that case 
when one goes from flat space to curved space, one simply replaces the
Minkowskian metric by the Riemannian metric (see, for example, Sect. 6.1, p. 122 of Trautman \textit{et al} \cite{Trautman1964}). In our case this is manifested by the Hubble law which replaces the
flat space expansion, thus replacing $\tau$ by $H_0^{-1}$ seems to be the 
proper thing to do. Accordingly, we get  
\begin{equation} \label{eqn:newredshiftformula4}
t=\frac{2H_0^{-1}}{1+\left(1+z\right)^2}.
\end{equation}
If we assume $H_0=70$ km/s-Mpc we finally obtain 
\begin{equation} \label{eqn:newredshiftformula5}
t \approx \frac{28}{1+\left(1+z\right)^2}\mbox{\rm Gyr}.
\end{equation}

In Fig. \ref{fig:fig2} we give the dependence of the cosmic time on the redshift for the 
entire range of $z$.  In order to visualize the dependence of cosmic time on redshift ($z$) in the
early Universe, equation (\ref{eqn:newredshiftformula5}) is shown in Fig. \ref{fig:fig3} on a logarithmic time axis.
Here $t$ is the time measured from the Big Bang.
\section{Concluding Remarks}
It is worth mentioning the physical assumptions behind the above mathematical 
formalism. First we have the principle of cosmological relativity according
to which the laws of physics are the same at all cosmic times. This is an
extension of Einstein's principle of relativity according to which the laws of
physics are the same in all coordinate systems moving with constant 
velocities. 

In cosmology the concept of time ($t=x/v$) replaces that of 
velocity ($v=x/t$) in ordinary special relativity. Second, we have the 
principle that the Big Bang time $\tau$ is always constant with the same 
numerical value, no matter at what cosmic time it is measured. This is obviously comparable to the assumption 
of the constancy of the speed of light $c$ in special relativity. 

Velocity in
expanding Universe is not absolute just as time is not absolute in special 
relativity. Velocity now depends on at what cosmic time an object (or a 
person) is located; the more backward in time, the slower velocity progresses, the more distances contract, and the heavier the object becomes. In the
limit that the cosmic time of a massive object approaches zero, velocities
and distances contract to nothing, and the object's energy becomes infinite. 

In Einstein's special relativity, as is well known, things depend on the velocity: The faster the object moves, the slower time progresses, the more 
distances contract, and the heavier the object becomes.

In this Note we have derived a simple formula, valid for all redshift values
in the Universe. The formula relates the cosmic time $t$ since the Big Bang,
for an earth observer at the present epoch, to the measured redshift $z$ of
light emitted at time $t$. It is hoped that the formula will be useful for
identifying objects at the early Universe since we can go back in time as far
as we desire but not to the Big Bang event at which the redshift becomes 
infinity.


\begin{thebibliography}{99}
\bibitem{Carmeli1995} Carmeli, M. 1995, Found. Phys., 25, 1029 
\bibitem{Carmeli1996} Carmeli, M. 1996, Found. Phys., 26, 413
\bibitem{Carmeli2002} Carmeli, M. 2002, \textit{Cosmological Special Relativity:The Large-Scale Structure
of Space, Time and Velocity, Second Edition}. (Singapore, World Scientific)
\bibitem{Cen1999} Cen, R. \& Ostriker, J.P. 1999, Ap. J, 514, 1-6
\bibitem{Landau1959} Landau, L. \& Lifshitz, E. 1959, \textit{The Classical Theory of Fields}, translated 
from the Russian by Morton Hamermesh, Second Printing (Edison-Wesley Publishing Company, Reading, Massachusets)
\bibitem{Schwarzschild2005} Schwarzschild, B. 2005, Physics Today, 58, March, 19-21 
\bibitem{Trautman1964} Trautman, A., Pirani, F.A.E. \& Bondi, H. 1964, Brandeis Summer Institute in
Theoretical Physics, Vol. 1: Lectures on General Relativity (Prentice-Hall, Inc. Englewood Cliffs, New Jersey)
\end{thebibliography}
\end{document}